\documentclass[a4paper,fleqn,usenatbib]{mnrasMOD}
\pdfoutput=1
\usepackage{newtxtext,newtxmath}
\usepackage[T1]{fontenc}
\usepackage{ae,aecompl}

\usepackage{graphicx}
\usepackage[font=small, font+=it, format=hang, margin={1cm, 1cm}]{caption}
\graphicspath{{figures/}}

\usepackage{amsmath}	% Advanced maths commands
\usepackage{amssymb}	% Extra maths symbols
\usepackage{multicol}
\usepackage{hyperref}
\usepackage{amsfonts}
\usepackage{empheq}
\usepackage{setspace}
\usepackage{color}
\usepackage{slashed}
 \usepackage{hyperref}
 \usepackage{mathrsfs}
 \usepackage{wasysym}
 \usepackage{booktabs}
\usepackage{float}
 \usepackage[british]{babel}
\usepackage{hyphenat}

\numberwithin{equation}{section}  

\usepackage{tensor}
\usepackage{tikz}

\usepackage{dcolumn}% Align table columns on decimal point
\usepackage{bm}% bold math
\usepackage{natbib}

\newcommand{\bea}{\begin{eqnarray}}
\newcommand{\eea}{\end{eqnarray}}
\newcommand{\be}{\begin{equation}}
\newcommand{\ee}{\end{equation}}

\def\p{\partial}

\title[Self-similar accretion in thin disks around near-extremal black holes]{Self-similar accretion in thin disks around near-extremal black holes}

\author[]{%
Geoffrey Comp\`ere,
Roberto Oliveri,
\vspace{0.2cm}
\\
Physique Th\'eorique et Math\'ematique and International Solvay Institutes, \\
Universit\'e Libre de Bruxelles, CP 231, B-1050 Bruxelles, Belgium\vspace{0.1cm}\\
Emails: \href{mailto: gcompere@ulb.ac.be}{gcompere@ulb.ac.be}, \href{mailto: roliveri@ulb.ac.be}{roliveri@ulb.ac.be}
}

\pubyear{2017}

\begin{document}
\label{firstpage}
\pagerange{\pageref{firstpage}--\pageref{lastpage}}
\maketitle


\begin{abstract}
Near-maximally spinning black holes display conformal symmetry in their near-horizon region, which is therefore the locus of critical phenomena. 
In this paper, we revisit the Novikov-Thorne accretion thin disk model and find a new self-similar radiation-dominated solution in the extremely high spin regime. Motivated by the self-consistency of the model, we require that matter flows at the sound speed at the innermost stable circular orbit (ISCO). We observe that, when the disk pressure is dominated by radiation at the ISCO, which occurs for the best-fitting Novikov-Thorne model of GRS 1915+105, the Shakura-Sunyaev viscosity parameter can be expressed in terms of the spin, mass accretion rate and radiative efficiency. We quantitatively describe how the exact thin disk solution approaches the self-similar solution in the vicinity of the ISCO and for increasing spins. 
\vspace{0.0cm}
\end{abstract}

%\tableofcontents
\begin{keywords} accretion, accretion disks --- black hole physics --- conformal symmetry --- radiation mechanisms: thermal --- X-rays: binaries 
\end{keywords}

%%%%%%%%%%%%%%%%%%%%%%%%%%%%%%%%%%%%%%%%%%%%%%%%%%

%%%%%%%%%%%%%%%%% BODY OF PAPER %%%%%%%%%%%%%%%%%%
\begingroup
\let\clearpage\relax
\tableofcontents
\endgroup
\vspace{-0.6cm}

\section{Introduction and summary}

The Novikov-Thorne geometrically thin and optically thick disk model (\cite{25334}) and its non-relativistic limit (\cite{1973A&A....24..337S}) are benchmarks of black hole accretion theory (see also \cite{25341,1975ApJ...200..187E,25363,1997ApJ...479..179A,2002ApJ...567..463L} for important corrections and refinements; for reviews, see \cite{Blaes:2002wz,2007A&ARv..15....1D,lrr-2013-1,Blaes:2013toa,2014ARA&A..52..529Y,2016ASSL..440....1L}). The model is analytical, based on falsifiable assumptions whose range of validity can be tested, and it only depends on four free parameters that include the phenomenological Shakura-Sunyaev $\alpha$ ``viscosity''. In particular, it predicts a blackbody thermal spectrum in the range $10^4-10^7$K which makes it suitable as a rough model of several classes of black hole binaries and luminous active galactic nuclei (see e.g. \cite{1999PASP..111....1K,McClintock:2013vwa}). 

One important boundary condition imposed in the original Novikov-Thorne model is that the torque vanishes at the ISCO, which leads to an inconsistency of the model (the fluid velocity diverges at the ISCO even though observables remain finite). This hypothesis has also been challenged by external considerations (\cite{1999ApJ...515L..73K,1999ApJ...522L..57G,2000ApJ...533L.115L,2010ApJ...711..959N}). Modelling physical boundary conditions at the ISCO is crucial in particular for highly spinning black holes in order to accurately calibrate their spin estimate (\cite{Li:2004aq}).  In \cite{Penna:2011rw}, a self-consistent boundary condition hereafter called \emph{``sonic-ISCO''} was proposed, which consists in equating the radial comoving fluid velocity at the ISCO with the sound speed. Two arguments were presented in its favour. First, in slim disk models around the Schwarzschild black hole analysed in \cite{Abramowicz:2010nk}, the sonic point asymptotes to the ISCO in the thin disk regime (where the accretion rate is sub-Eddington), independently of $\alpha$. Secondly, the boundary condition implies that advection terms are negligible with respect to the stresses at the ISCO in the energy balance, in the limit where the disk height is negligible with respect to $\alpha$. We will check that the latter assumption is self-consistent for typical stellar-mass black holes and luminous AGN parameters. A crucial consistency condition of thin disk is therefore satisfied. As we will discuss, the boundary condition also implies that the specific internal energy is negligible, which automatically enforces another hypothesis of the Novikov-Thorne model. In addition, we will show that the sonic-ISCO boundary condition allows for a well-defined near-horizon near-extremal scaling behavior, in contrast to the no-torque boundary condition. Our first objective is to derive the complete thin disk model with the sonic-ISCO boundary condition from first principles. We took particular care to correct various algebraic errors and typos found in the literature. 

A new feature emerges from our analysis. For highly spinning stellar-mass black holes, gas pressure at the ISCO becomes negligible with respect to radiation pressure. The sonic-ISCO boundary condition then implies that the accretion rate is not a free parameter of the model. Instead, the disk height at the ISCO is the fourth independent parameter. It allows us to infer a best-fitting value for the Shakura-Sunyaev parameter $\alpha$ in terms of the total luminosity, radiative efficiency and spin of the source. For example in the case of the X-ray binary source GRS 1915+105, recent estimates of the mass and spin, using the Very Long Baseline Array, give $M \sim 12 \pm 2$ M$_{\astrosun}$ and $J/M^2=0.98 \pm 0.01$  (\cite{Reid:2014ywa}). Also, the best-fitting model of the \emph{NuSTAR} observation of GRS 1915+105 in the plateau state gives a disk luminosity at $23\% \pm 4 \%$ of the Eddington rate  (\cite{2041-8205-775-2-L45}). Taking into account the torque contribution to the radiative efficiency of thin accretion disks (\cite{2002ApJ...567..463L}), we will infer that the best-fitting Novikov-Thorne model is a disk dominated by radiation around the ISCO with $\alpha =0.43$ and $\dot M =1.6 \times 10^{18} g\ sec^{-1}$.

It is well established that radiation-dominated phases in thin disks are both thermally and viscously unstable (\cite{1974ApJ...187L...1L,1976MNRAS.175..613S}). Without a stabilization mechanism, such disk models are not expected to occur as steady flows in nature or be visible in numerical simulations (see the thermal instability in \cite{Mishra:2016yyp}). Instead, the radiation pressure instability has been argued to drive the variability of radiation-dominated objects such as GRS 1915+105 (\cite{1538-4357-542-1-L33}). The thermal and viscous instabilities are the main open issues to be understood in the Novikov-Thorne model. In this paper, we will further explore the radiation-dominated phase with the sonic-ISCO boundary condition in the extremely high spin regime, where new features arise. 

According to the cosmic censorship conjecture, the spin of the Kerr black hole is bounded by its value at extremality.  Extreme spinning black holes admit a near-horizon region with global conformal symmetry $SO(2,1)$ (\cite{Bardeen:1999px}) and probe fields can therefore be described using the language of conformal representation theory (\cite{Porfyriadis:2014fja,Lupsasca:2014pfa,Zhang:2014pla,Lupsasca:2014hua,Li:2014bta,Compere:2015pja}) or critical phenomena (\cite{Gralla:2016jfc}). Approaching the extremality bound by realistic accreting processes is, however, limited by the absorption cross-section of retrograde photons as shown by \cite{1974ApJ...191..507T}, leading to the bound $J/M^2 <99.8\%$. This bound is commonly accepted (see e.g. recent work by \cite{Kesden:2009ds}), even though it can be lowered by magnetic fields (\cite{Gammie:2003qi}) %or gravitational wave emission (\cite{Gralla:2015rpa}) 
or instead challenged (\cite{2011A&A...532A..41S}). More fundamentally, for accretion disk models where the inner edge  approaches the marginally bound orbit instead of the innermost stable orbit, the capture of retrograde photons asymptotes to zero, which allows spinning black hole to further approach extremality (\cite{1980AcA....30...35A}). Such scenarios have been shown to occur in slim disk models (\cite{2011A&A...532A..41S}). It is therefore worthwhile to explore which specific signatures may arise in the extremely high spin regime, where conformal symmetry is only slightly broken. This programme has already been carried out for gravitational wave signatures (\cite{Hadar:2014dpa,Hadar:2015xpa,Gralla:2016qfw}) and we initiate this programme for accretion disks in the simplest case of thin disk models. (for earlier related work see also \cite{2009arXiv0909.2041G}). 

The paper is organized as follows. In Sec. \ref{sec:cr}, we first show how the emergent conformal symmetry constraints finite observables in the near-extremal near-horizon limit of Kerr spacetime, which completes the analysis of \cite{Gralla:2016jfc}.
Sec. \ref{sec:disk} is devoted to revisit the thin disk equations and the main assumptions of the Novikov-Thorne model. We impose the sonic-ISCO boundary condition (\cite{Penna:2011rw}), and we explain its physical implications. In Sec. \ref{sec:features}, we analyse the features of the Novikov-Thorne model and we construct piece-wisely the global solution. We focus on parameters for two notable cases of interest: stellar-mass and supermassive black holes. Moreover, we study the phase diagram of thin accretion disks in either cases in terms of the spin parameter. Remarkably, thin accretion disks  reach a critical spin beyond which a phase transition occurs, close to the ISCO, from gas pressure dominated to radiation pressure dominated. This happens in the high spin regime. As an example, we consider the Novikov-Thorne model parameters to be those of the best-fitting values of GRS1915+105. Finally, in Sec. \ref{NHEK sol}, we present a first explicit example of critical (self-similar) disk in the near-horizon region of near-extremal Kerr. In Appendix \ref{appKerr}, the reader can find all the necessary properties of the Kerr spacetime and the main definitions we used throughout the paper. Appendix \ref{Solutions} contains the local solutions to the thin disk equations.

%
% Modeling is pressing, given the abundance of data to explain today, e.g. the need to model a large class of X-ray binaries with low variability of \cite{Gierlinski:2003dg}, and tomorrow, e.g. explaining the black hole silhouette \cite{Straub:2012uv} which will be visible in the future by the Event Horizon Telescope. 

\section{Critical phenomena around near-extremal black holes}
\label{sec:cr}

It was originally discovered by \cite{Bardeen:1999px} that in the extreme spinning limit 
\bea
\frac{J}{M^2}=\sqrt{1-\sigma^2},\quad \sigma \ll 1,
\eea
a scale invariant spacetime region emerges in the near-horizon region of the Kerr black hole. More precisely, starting with the extreme ($\sigma = 0$) Kerr black hole in Boyer-Linquist (BL) coordinates $(t,r,\theta,\phi)$, one can take the near-horizon corotating limit by defining new coordinates 
$(T, R, \theta, \Phi)$ as
\be\label{NHEKs}
T = \frac{\lambda t}{2 M}, \qquad R = \frac{r - M}{\lambda M}, \qquad \Phi = \phi - \frac{t}{2M},
\ee
and let $\lambda \rightarrow 0$ with $T$, $R$ and $\Phi$ fixed. The asymptotically flat region then completely decouples and the resulting near-horizon extreme Kerr (NHEK) geometry is geodesically complete and given by
\be \label{NHEK}
ds^2 = 2M^2 \Gamma\left[-R^2dT^2 +\frac{dR^2}{R^2} + d\theta^2 + \gamma^{2}\left(d\Phi + RdT\right)^2\right],
\ee
where $\Gamma(\theta) = \left(1+\cos^2(\theta)\right)/2$ and $\gamma(\theta) = \sin(\theta)/\Gamma(\theta)$. 

Since it is obtained as a scaling limit of Kerr, it obeys the vacuum Einstein's equations and it admits enhanced scaling symmetry along the Killing vector $H_0 \equiv T \p_T - R \p_R$. It turns out that the exact symmetry is further enhanced to the conformal group times the abelian group of azimuthal rotations, $SO(2,1)\times U(1)$. 

Conformal symmetry also appears for infinitesimal deviations from extremality (\cite{Amsel:2009ev,Bredberg:2009pv}), which implies that near-extremal black holes admit critical phenomena in their near-horizon region.  In order to make that statement precise in the presence of matter accretion, it is fundamental to control the location of the ISCO.\footnote{Local conformal symmetry also arises (\cite{Guica:2008mu}) and plays an important role in understanding the quantum physics of black holes (see the reviews \cite{Bredberg:2011hp,2012arXiv1203.3561C}). Conformal symmetry also appears for specific probes away from extremality (\cite{Castro:2010fd}). We will, however, not use these additional structures for the present considerations.}

It has been already noticed in the original paper of \cite{Bardeen:1972fi} (see their Fig. 2) that the BL coordinate radius of the ISCO ($r_{0}$), as well as photon and marginally bound orbits, asymptotically merges with the BL coordinate radius of the horizon $r_{+}$ towards extremality,
\be
\frac{r_{+}}{M} = 1 + \sigma, \qquad \frac{r_{0}}{M} = 1 + 2^{1/3}\sigma^{2/3}+\mathcal{O}\left(\sigma^{4/3}\right) .
\ee
Therefore, in order to obtain a finite coordinate location for the ISCO in the near-horizon limit, it is necessary to scale the near-extremality parameter as $\sigma = \bar\sigma \lambda^{3/2}$, where $\bar{\sigma}$ is fixed, as described in \cite{Gralla:2015rpa}. The $\lambda \rightarrow 0$ limit still gives the NHEK spacetime in Eq. \eqref{NHEK} and the ISCO is now located at the finite radial position $R=R_{0} \equiv 2^{1/3} \bar{\sigma}^{2/3}$. 

The near-horizon near-extremal limit above is invariant under the scaling $\lambda \rightarrow c \lambda$ for arbitrary $c$. Therefore, all physical quantities, which are finite in this limit, will be invariant under a simultaneous rescaling of $R \rightarrow c R$, $T \rightarrow T/c$ and $R_{0} \rightarrow c R_{0}$. In other words, finite observables in the above limit are zero eigenfunctions of the total dilatation operator $T \p_T - R \p_R - R_{0}\p_{R_{0}}$. In particular, stationary observables $O$ ought to be of the form $\left(R_{0}/R\right)^{{\mathfrak h}}$, where ${\mathfrak h}$ is the critical exponent, which equals the conformal weight of the observable in the near-horizon region
\bea
\mathcal L_{H_0} O = {\mathfrak h}\, O.
\eea 
Thanks to an additional length-scale, namely $R_0$, near extremal configurations allow a richer variety of critical exponents as compared with the extremal case discussed in \cite{Gralla:2016jfc}.

In Sec. \ref{NHEK sol}, we will provide an explicit example of such a critical behavior within the Novikov-Thorne accretion disk model with critical exponents
\bea\label{cw}
 {}[ u^R]=[h]= [F]=2,\qquad [T]=[p]=0,\qquad [ \rho ]=-4.
\eea
Here, $u^R$ is the near-horizon radial velocity, $h$ the disk opening angle, $F$ the vertical energy flux, $T$ the temperature, $p$ the pressure and $\rho$ the rest-mass density.

\section{Thin disk equations} \label{sec:disk}
In this section, we revisit the accretion thin disk model around Kerr spacetime, originally presented in \cite{25334} and \cite{25341}, but instead of imposing the \emph{``no-torque''} boundary condition at the ISCO, we will impose the \emph{``sonic-ISCO''} boundary condition introduced in \cite{Penna:2011rw}. The familiar reader might skip sub-sections \ref{sec:1}-\ref{sec:3} and directly reach sub-section \ref{ISCOBC}, where new features arise. 

\subsection{Fundamental equations}\label{sec:1}
We assume a single component relativistic viscous fluid, flowing along the four-velocity $u^{\mu}$ in the fixed Kerr background. 
The stress-tensor can then be split uniquely with respect to $u^\mu$ as 
\be
T^{\mu\nu} = \rho (1+\Pi) u^{\mu}u^{\nu} + p h^{\mu\nu} + S^{\mu\nu} + u^{\mu}q^{\nu} + u^{\nu}q^{\mu},
\ee
where $\rho$ is the rest-mass density, $\Pi$ the specific internal energy, $p$ the total isotropic pressure, $h_{\mu\nu}= u_{\mu}u_{\nu} + g_{\mu\nu}$ the projector into the local rest frame (LRF), $S^{\mu\nu}$ the symmetric, transverse and traceless anisotropic stress-tensor that takes into account viscous stresses, and $q^{\mu}$ the transverse energy flux. All these quantities are relative to the LRF and all indices are raised with $g^{\mu\nu}$. 
The disk is assumed not to self-irradiate. Magnetic fields are ignored except for their contribution to the viscous stresses. Neutrinos and dark matter are ignored.

The fundamental equations governing the dynamics of the accretion disk are
\begin{equation} \label{fundamental eqs}
0 = \left(\rho u^{\mu}\right)_{; \mu}, \qquad 0 = h_{\mu\sigma}T^{\sigma\nu}_{\;\;\;\;\; ; \nu}, \qquad
0 = u_{\mu}T^{\mu\nu}_{\;\;\;\; ; \nu},
\end{equation}
respectively, the rest-mass conservation law, the relativistic Navier-Stokes equations and the energy conservation equation. These fundamental equations must be supplemented by the equation of state, the radiative energy transport law and prescriptions about the nature of the viscous effects and opacity.  Conservation of the rest-mass along the fluid four-velocity is valid for energies much below $2m_b$, where $m_b$ is the rest-mass of the baryon species, or equivalently, for temperatures much below $10^{10}$ K for electrons and $10^{13}$ K for protons/neutrons. We will see that the last hypothesis is self-consistent after presenting the solution to the model. 

Assuming thermal equilibrium and the existence of an equation of state of the form $\Pi=\Pi(p,v)$ where $v=1/\rho$ is the specific volume of the fluid, one can define the temperature $T$ and specific entropy $s$ by the first law $T ds = d\Pi +p dv$. The energy conservation equation might then be written in terms of local entropy production (\cite{Ellis:1971pg})
\be
\rho T u^{\mu}s_{,\mu} = - \left( S^{\mu\nu}\sigma_{\mu\nu} + q^{\mu}a_{\mu} + q^{\mu}_{\;\; ;\mu}\right),
\ee
where $\sigma_{\mu\nu} = h^{\alpha}_{\;\; \mu}h^{\beta}_{\;\; \nu}u_{(\alpha ; \beta)} - (1/3)h_{\mu\nu}u^{\alpha}_{\;\; ;\alpha}$ is the shear tensor and $a^{\mu} = u^{\nu}u^{\mu}_{\;\; ;\nu}$ is the acceleration. The four-entropy flux is defined as $S^{\mu} = s \rho u^{\mu} + q^{\mu}/T$ and the local form of the second law of thermodynamics $S^{\mu}_{\;\; ;\mu} \geq 0$ implies the two constitutive equations 
\be\label{consrel}
S^{\mu\nu} = - \eta \sigma^{\mu\nu}, \qquad q^{\mu} = -\lambda h^{\mu\nu}\left(T_{,\nu} + T a_{\nu}\right)
\ee
with $\eta, \lambda \geq 0$ being the viscosity and heat conduction coefficients, respectively.

\subsection{Thin disk approximation}\label{sec:2}
The fundamental equations \eqref{fundamental eqs} require additional simplifying assumptions in order to construct analytical accretion disk models. The original thin disk model provides a set of working assumptions, carefully listed in \cite{25341}, to transform the full system of partial differential equations to an algebraic non-linear system of equations and obtain local solutions in analytical form.

In what follows, we will use cylindrical BL coordinates $(t,r,z,\phi)$ where $z=r \cos(\theta)$ (see Appendix \ref{appKerr} for details and notation). We assume stationarity and axisymmetry. The half-thickness $H$ of the disk is defined as the vertical distance between its upper surface and the equatorial plane at $z=0$. A geometrical thin disk has an opening angle $h(r) \equiv H/r \ll 1$. We suppose that all radiation is vertical, $q^\mu \sim \delta^\mu_z$. We are only interested in vertically integrated quantities between $z=-H$ and $z=+H$. For example the surface density of the disk is defined as
\be \label{surface density}
\Sigma(r) \equiv \int_{-H}^{H} \rho(r, z) dz = 2 \rho H.
\ee
In order to directly obtain the equations for vertically integrated quantities from Eqs. \eqref{fundamental eqs} as the leading order of a Taylor expansion at the equator, one can simply ignore all $z$ dependence of all physical quantities, except for the pressure $p$ and radiation flux $q^z$, which can be assumed to be (see discussion in \cite{1997ApJ...479..179A}):
\bea
p(r,z) =p(r) \left(1-\frac{z^2}{H^2}\right),\qquad q^z(r,z) = F(r) \frac{z}{H} \;\;\; (z \leq H). 
\eea
The function $F$ is then the radiation flux emitted from the upper or lower side of the disk. The orbital motion of the fluid is taken to follow nearly circular equatorial geodesics with a small radial (non-geodesic) component $u^r$ produced by viscous stresses and responsible for accretion on to the black hole. 

Since the shear tensor $\sigma_{\hat \mu \hat \nu}$ only admits the non-vanishing component $\sigma_{\hat r\hat \phi} = \sigma_{\hat \phi \hat r} < 0$ in the local rest frame (see Eq. \eqref{shear tensor}), and given the first constitutive equation \eqref{consrel}, one can define the vertically integrated shear tensor as 
\be \label{int shear tensor}
W(r) \equiv \int_{-H}^{H} S_{\hat{r}\hat{\phi}}(r, z) dz = 2 S_{\hat{r}\hat{\phi}} H. 
\ee

We assume the $\alpha$-viscosity prescription of \cite{1973A&A....24..337S} 
\bea
S_{\hat{r}\hat{\phi}} = \alpha p,
\eea
where $\alpha$ is a free parameter and $p$ is total pressure. \footnote{Some specific MHD turbulent disks might be modelled by this prescription (\cite{1999ApJ...521..650B}). The modification of $\alpha p$ to $\alpha p_{gas}$ was argued in \cite{1974ApJ...187L...1L} and further developed in \cite{1977A&A....59..111B} and \cite{1981ApJ...247...19S}. Such models, later called $\beta p$ models, do not suffer from thermal instabilities. However, MHD simulations do not conclude on their validity (\cite{Hirose:2008hi,2009PASJ...61L...7O,Ross:2015gga}). More elaborated prescriptions have been developed by \cite{2003MNRAS.340..969O} and \cite{PhysRevLett.97.221103}.} %\cite{2008MNRAS.383..683P}

Two further assumptions were used in the original analysis of \cite{25334}, namely negligible specific internal energy density $\Pi = 0$ and negligible advection. We will see that these hypotheses are obeyed as a result of the sonic-ISCO boundary conditions and $ h \ll \alpha $, which will be obeyed in turn by explicit check on the solutions.

\subsection{Thin disk equations}\label{sec:3}

The dynamics in the radial direction is governed by the rest-mass, energy and angular momentum conservation laws. The radial Navier-Stokes equation is trivially satisfied with the additional assumptions that velocity and pressure gradients are negligible. Standard manipulations of Eqs. \eqref{fundamental eqs} lead to 
\begin{subequations} 
\begin{align}
\dot{M} &= -2 \pi r \Sigma u^{r}, \label{re1} \\
F &= - \sigma_{\hat{r}\hat{\phi}} W, \label{re2} \\
-4 \pi r \frac{(E - \Omega L)^2}{\Omega_{,r}} \frac{F}{\dot{M}} &= \int_{r_0}^{r}(E - \Omega L)L_{,r^{\prime}} dr^{\prime} + M \mathcal P_0. \label{re3} 
\end{align}
\end{subequations}
Equation \eqref{re1} is the conservation of rest mass, Eq. \eqref{re2} conservation of energy and Eq. \eqref{re3} is a combination of energy and angular momentum conservation laws.
The constant of integration $\dot{M}$ is the accretion rate, while $E$, $L$, $\Omega$ and $\sigma_{\hat{r}\hat{\phi}}$ are kinematic quantities of circular equatorial geodesics (see Appendix \ref{CEGs} for their expressions). 
The right-hand side of Eq. \eqref{re3} is nothing else than $M \mathcal{P}$ defined in \eqref{P}. 
The dimensionless integration constant $\mathcal P_0$ is fixed by the boundary conditions, which are discussed in sub-section \ref{ISCOBC}.

The vertical Navier-Stokes equation describes the pressure balance along the cylindrical vertical coordinate $z$ and reads as
\be   \label{vertical equation1}
\frac{2p}{\rho} = h^2 \frac{\mathcal{L}_{\star}^2}{r^2}
\ee
as derived in \cite{1997ApJ...479..179A}\footnote{Note the factor 2 typo in Eq. (B12) of \cite{Penna:2011rw}.}. The total pressure is the sum of the radiation pressure and the gas pressure 
\begin{subequations}\label{EOS}
\begin{align}
&p = p^{(gas)} + p^{(rad)} ,\\
&p^{(gas)} = \frac{k_B \rho}{m_p} T,\\
&p^{(rad)} =\frac{1}{3} b T^4,
\end{align}
\end{subequations}
%$b \equiv \frac{8 \pi^5}{15} \frac{k_B^4}{c^3 h_P^3}$%
where $b = 4\sigma_{SB}/c$ is the radiation constant density, $k_B$ Boltzmann's constant, $\sigma_{SB}$ Stefan-Boltzmann's constant and $m_p$ the rest-mass of the proton.

We impose the energy transport law
\be \label{vertical equation2}
b T^4 =\bar{\kappa} \Sigma F
\ee
where $\bar\kappa$ is the optical opacity of the disk
\begin{subequations} \label{opacity prescriptions}
\begin{align}
\bar{\kappa} &= \bar{\kappa}_{ff} + \bar{\kappa}_{es},\\
\bar{\kappa}_{ff} &= \left(0.64 \times 10^{23} cm^2 g^{-1}\right)\left(\frac{\rho}{g/cm^3}\right)\left(\frac{T}{K}\right)^{-7/2},\\
\bar{\kappa}_{es} &= 0.40 \; cm^2 g^{-1} ,
\end{align}
\end{subequations}
originating from free-free (ff) absorption and electron scattering (es).

\subsection{Sonic-ISCO boundary condition}
\label{ISCOBC}

A physical system is determined by its equations of motion and its boundary conditions.  
In the original analysis of \cite{25334}, it was assumed that there is no torque at the ISCO (located at $r=r_0$), which is equivalent to assume that there is no radiation at that point,
\be
F(r_0) = 0,
\ee
which is also equivalent to fixing the integration constant $\mathcal P_0 =0$  in Eq. \eqref{re3}.

Following \cite{Penna:2011rw}, we impose instead that the (purely radial) fluid velocity in the frame corotating with the geodesic flow equals (minus) the sound speed at the ISCO, 
\be
c_s (r_0) =-u^{\hat r}(r_0). \label{sound}
\ee
This boundary condition has three important advantages that we will describe in what follows. 

So far, we can rewrite the energy conservation equation as
\be
Q_{diss} = Q_{cool}+Q_{adv},
\ee
where $Q_{diss} = - S^{\mu\nu} \sigma_{\mu\nu}$ is the dissipation function, $Q_{cool} = q^{\mu}_{\;\; ;\mu}$ is the cooling function and $Q_{adv} = \rho Tu^{\mu}s_{,\mu}$ is the advection function that takes into account the rate of change of the specific entropy along the four-velocity. Now, as shown in \cite{Penna:2011rw}, the boundary condition \eqref{sound} implies the following scaling relations:  $Q_{adv} \sim h^2$, $Q_{cool}\sim Q_{diss} \sim \alpha h$; so if $h \ll \alpha$, advection can indeed be neglected. 

Thin disk models usually assume that the specific internal energy density is negligible $\Pi = 0$. This hypothesis is justified if the sound speed is nonrelativistic, $c_s \ll 1$. Due to the gravitational potential, the sound speed is highest in the near-horizon region of the disk, where we will impose the boundary condition \eqref{sound}. The hypothesis $c_s \ll 1$ will therefore be obeyed as long as $|u^{\hat r}| \ll 1$ at the ISCO, which is already part of the hypotheses since we assumed that the fluid follows nearly circular geodesics. We will check that the solutions indeed obey $|u^{\hat r}| \ll 1$ and $c_s \ll 1$. 

Finally, with the no-torque boundary condition, the disk model has no regular limit at the ISCO as we will review below, while the boundary condition \eqref{sound} regularizes the model. 

Let us now fix the remaining integration constant $\mathcal P_0$ for the sonic-ISCO boundary condition. It will be fixed as a function of the disk height at the ISCO, $h_0$, as follows. The integral in the right-hand side of \eqref{re3} is zero when evaluated at the ISCO. Thus, Eq. \eqref{re3} reads as
\be \label{c1}
\left(\frac{F}{\dot{M}}\right)_0 = - \frac{M \Omega_{,r}|_0 \mathcal P_0}{4\pi r_0(E_0 - \Omega_0 L_0)^2}.
\ee
On the other hand, dividing the first two equations \eqref{re1} and \eqref{re2}, we get
\be \label{c2}
\left(\frac{F}{\dot{M}}\right)_0 = \left(\frac{\sigma_{\hat{r}\hat{\phi}} W}{2 \pi r \Sigma u^{r}}\right)_0 =  - \frac{1}{\sqrt{2}} \frac{A_0}{4\pi r^{4}_0 \Delta_0^{1/2}}\alpha h_0\gamma^2_0\mathcal{L}_{\star,0}\Omega_{,r}|_0.
\ee
In the second step, we have expressed the radial velocity component in the LRF, $u^{r} = (\Delta^{1/2}/ r) u^{\hat{r}}$, we have substituted the definitions of $\Sigma$ and $W$, we have used the Shakura-Sunyaev prescription and assumed the sonic-ISCO boundary condition \eqref{sound}, $u^{\hat{r}}_{0} = -\sqrt{p_0 /\rho_0}$. Equating \eqref{c1} and \eqref{c2}, we find
\begin{align} \label{C}
\mathcal P_0 &= \frac{1}{\sqrt{2}} \frac{\alpha h_0}{M} \frac{A_0}{r^3_0 \Delta^{1/2}_0} \gamma^2_0 (E_0 - \Omega_0 L_0)^2 \mathcal{L}_{\star,0}  \nonumber\\
& = \frac{1}{\sqrt{2}}  \alpha h_0 x_0 \mathcal{D}^{1/2}_{0} \mathcal{R}^{1/2}_{0},
\end{align}
after some algebra involving the functions defined in Appendix \ref{appKerr}. Our final formula for $M \mathcal P_0$ disagrees with the constant $C$ derived in \cite{Penna:2011rw}, which is easily seen to be incorrect since it has the wrong dimension of length. 

The physical meaning of the integration constant $\mathcal{P}_{0}$ is to introduce a torque at the ISCO. More precisely, the torque might be derived by comparison of Eq. \eqref{re3} and Eq. (12) of \cite{2002ApJ...567..463L}. The torque $g_{0}$ is then
\be \label{torque}
g_{0} = \frac{M \dot{M} \mathcal{P}_{0}}{E_{0} - \Omega_{0}L_{0}}.
\ee
The total energy radiated per unit time as measured by an observer at infinity is therefore given by both accretion and torque contribution (\cite{2002ApJ...567..463L})
\be
\mathcal{L}_{tot} =  \eta_{0} \dot{M} + g_{0} \Omega_{0} \equiv \eta \dot{M},
\ee
where $\eta_{0}  = 1 - E_{0} $ is the specific conserved energy of a particle orbiting along the ISCO \eqref{kin quantities1} and $\eta$ is the radiative efficiency of the disk 
\be \label{radiative efficiency}
\eta = \eta_{0} + \frac{g_{0}\Omega_{0}}{\dot{M}} = \eta_0\left(\frac{a}{M}\right) + \alpha h_0 g\left(\frac{a}{M}\right),
%= \eta_{0}\left(1 + \frac{M\mathcal{P}_{0}}{\eta_0}\frac{\Omega_{0}}{E_{0} - \Omega_{0}L_{0}}\right).
\ee
where $g\left(\frac{a}{M}\right)=2^{-1/2}M\Omega_0 (E_{0} - \Omega_{0}L_{0})^{-1}x_0 \sqrt{\mathcal D_0 \mathcal R_0} = 2^{-1/2}x_{0}^{-2}\mathcal{C}_{0}^{-1/2}\sqrt{\mathcal{D}_0 \mathcal{R}_0}$. Here, we are ignoring the capture of radiation from the hole, which decreases the efficiency for high spin (\cite{1974ApJ...191..507T}).

\section{Features of the general solution} \label{sec:features}

We introduce the dimensionless mass and mass accretion rate
\be
M_{\star} \equiv \frac{M}{3 \mbox{M}_{\astrosun}}, \qquad \dot{M}_\star \equiv \frac{\dot{M}}{10^{17} g\; sec^{-1}}
\ee 
where M$_{\astrosun}$ is the mass of the Sun. 

The global solution can be approximated by a piece-wise construction of three local solutions described in Appendix \ref{Solutions}, which are patched according to their range of validity. The qualitative features of the global solution depend upon the region that dominates at the ISCO. There are three possibilities depending which of the three relevant local solution is valid around the ISCO. 

\subsection{Gas-pressure-dominated ISCO}

In the ``standard'' first two cases, the ISCO lies in the region dominated by gas pressure, either Region [Gas-es] or Region [Gas-ff] of Appendix \ref{Solutions}. The four free parameters of the model are $(M,a,\dot M,\alpha)$ and the disk height at the ISCO, $h_0$, is fixed. Indeed, if the ISCO lies in Region [Gas-es], we evaluate \eqref{h gases} at the ISCO using Eq. \eqref{C}. The result is 
\be \label{h0 gases}
h_0 = \left(1.8 \times 10^{-3}\right)\left( \alpha^{1/8} M_{\star}^{-3/8} \dot{M}_{\star}^{1/4}\right) x_0^{1/8}\mathcal{C}_0^{-1/8}\mathcal{R}_0^{-1/2}.
\ee
If the ISCO lies in Region [Gas-ff] instead, we evaluate \eqref{hB3} at the ISCO using Eq. \eqref{C}. The result is 
\be \label{h0B3}
h_0 = \left(1.3 \times 10^{-3}\right)\left( \alpha^{1/17} M_{\star}^{-5/17} \dot{M}_{\star}^{3/17}\right) x_0^{5/17}\mathcal{C}_0^{-1/17}\mathcal{D}_0^{-1/34}\mathcal{R}_0^{-8/17}.
\ee
We checked that the hypothesis $h_0 \ll \alpha$ is obeyed at the ISCO in both cases \eqref{h0 gases} and \eqref{h0B3} for the range $\alpha \sim 0.01-1$ and $a/M \sim 0 - 0.999$ and either $(M_{\star},\dot{M}_\star)\sim (1, 1)$ or $(M_{\star},\dot{M}_\star)\sim (10^{7}, 10^{5})$.

\begin{figure}
\centering
\includegraphics[width=8.5cm]{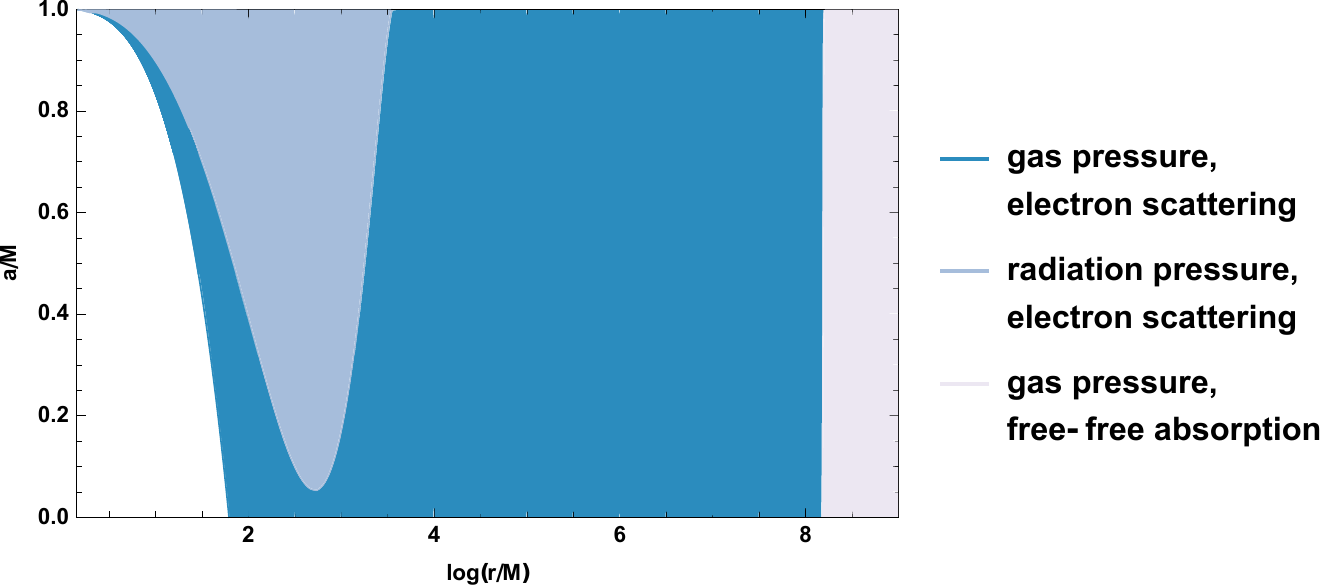}
\caption{Disk regions for stellar-mass black holes with $(M_{\star},\dot{M}_\star)= (1, 1)$ and $\alpha = 0.2$ in the spin range $0 \leq a \leq 0.999 M$.}
\label{regions}
\end{figure}

\begin{figure}
\centering
\includegraphics[width=8.5cm]{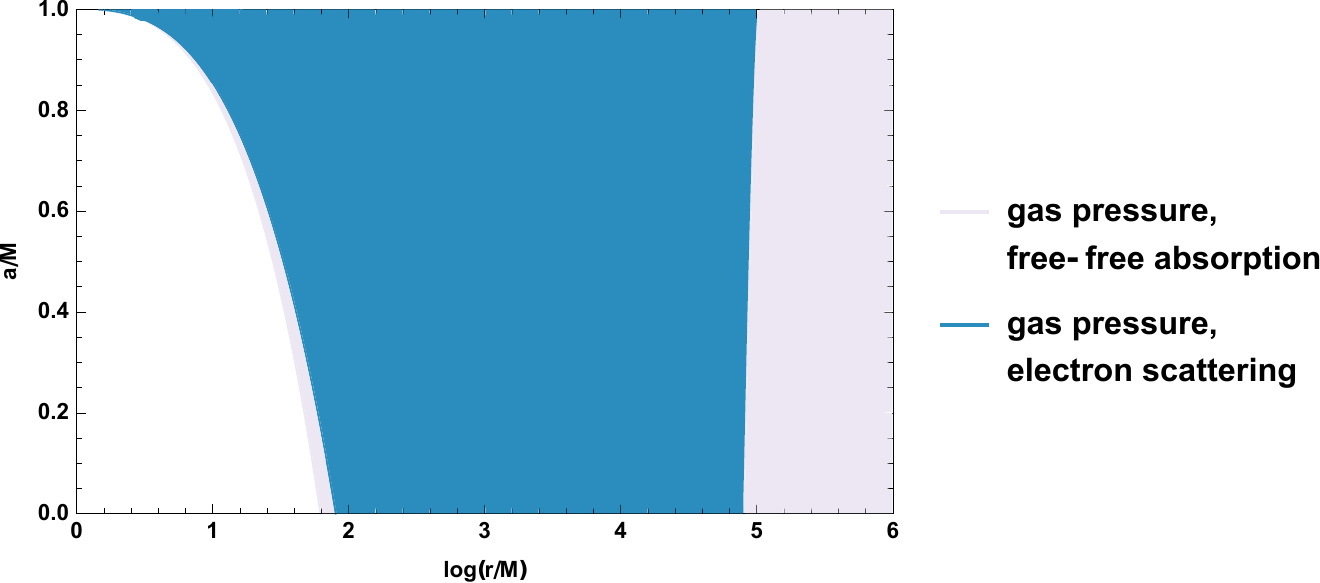}
\caption{Disk regions for supermassive black holes with $(M_{\star},\dot{M}_\star)= (10^{7}, 10^{5})$ and $\alpha = 0.2$ in the spin range $0 \leq a \leq 0.999 M$.}
\label{regions2}
\end{figure}

As explicit examples, the disk regions and their transitions are plotted in Figs \ref{regions} and \ref{regions2} for the spin range $0 \leq a \leq 0.999 M$ assuming $\alpha = 0.2$ for either $(M_{\star},\dot{M}_\star)=(1, 1)$ (modelling a stellar-mass black hole) and $(M_{\star},\dot{M}_\star)= (10^{7}, 10^{5})$ (modelling an AGN).

In Fig. \ref{regions}, the first region in which the ISCO is located is called the \emph{edge region}. There, the gas pressure overwhelms the radiation pressure and the opacity due to electron scattering is dominant over the free-free absorption. The disk height is given by Eq. \eqref{h0 gases}. The transition to the \emph{inner region} occurs when radiation pressure starts becoming predominant over the gas pressure. A new transition occurs when the gas pressure becomes dominant again over radiation pressure and the resulting region is called the \emph{middle region}. For very low spins, the inner region is absent and the edge and middle regions merge.  The \emph{outer region} is still gas-pressure-dominated, but the main mechanism responsible for the opacity is the free-free absorption.  We checked that the height satisfies $h \ll \alpha = 0.2$ for $r \ll 10^{13}M$ where the model breaks down for other reasons (because self-gravitation is not negligible). The temperature of the accretion disk is hotter in the region near the ISCO, and in general the temperature is higher for faster spin. By evaluating the temperature at the ISCO for $a=0.999M$, we found $T \sim 10^7 $K. This implies that the assumption of conservation of  mass is valid. We also checked that the sound speed is negligible with respect to light speed, $c_s =\sqrt{p/\rho}\ll c$, in the entire disk. 

The original Novikov-Thorne model did not contain the edge region, but it should contain it by consistency: for standard spins, the radiation pressure in the inner region goes to zero at the ISCO and therefore the gas pressure needs to dominate at low enough radius. Yet, if one assumes the no-torque boundary condition, the Novikov-Thorne model is singular at the ISCO in the edge region. The reason is readily seen because in the [Gas-es] solution  \eqref{sol gases} the function $\mathcal{P}$ defined in \eqref{P} vanishes at the ISCO if $\mathcal P_0=0$, and the radial velocity is then divergent at the ISCO. The non-zero torque introduced by the sonic-ISCO boundary condition allows us to regulate the model as claimed earlier.

In Fig. \ref{regions2}, the disk is always dominated by gas pressure, but the dominant contribution to opacity varies with radius. In the \emph{edge region}, where the ISCO lies, the free-free absorptions are dominant. The disk height is therefore given by Eq. \eqref{h0B3}. The \emph{middle region} is dominated by electron scattering. The \emph{outer region} is again dominated by free-free absorption. The assumption $h \ll \alpha = 0.2$ is obeyed for $r \ll 10^{21} M$ where the model breaks down. The accretion disk for supermassive black holes is colder with respect to the stellar-mass black holes; it never exceeds $T \sim 10^4$K and the conservation of mass is obeyed. We also checked that the sound speed is negligible with respect to light speed, $c_s \ll c$, in the entire disk. 

\subsection{Radiation-pressure-dominated ISCO}

Let us now discuss the configurations where the ISCO lies in the region dominated by radiation pressure and electron scattering (Region [Rad-es] of Appendix \ref{Solutions}). This scenario happens for very high spins as we will discuss below. A new feature arises as a result of the sonic-ISCO boundary condition: a constraint relates the accretion rate $\dot{M}$, the mass, spin and $\alpha$ parameter, while the opening angle at the ISCO is unconstrained. 

Indeed, the disk height \eqref{h rades} in geometric units is given by
\be
h =  \frac{\bar{\kappa}_{es}}{2\pi}\frac{\dot{M}}{M} x^{-3}\mathcal{C}^{-1}\mathcal{R}^{-1}\mathcal{P}.
\ee
At the ISCO,  $\mathcal{P}$ is given by $\mathcal P_0$ (see Eq. \eqref{P}) that can be evaluated using the expression for the sonic-ISCO boundary condition in Eq. \eqref{C}. Therefore, $h_0$ appears linearly in both sides and we are left with a constraint among the parameters of the model given by
\be \label{constraint rades}
\frac{\alpha \bar{\kappa}_{es}}{4\pi} \frac{\dot{M}}{M} = \frac{1}{\sqrt{2}} x_0^{2} \mathcal{C}_0 \mathcal{D}_0^{-1/2}  \mathcal{R}_0^{1/2} \equiv f \left(\frac{a}{M}\right)
\ee
The function $f\left(\frac{a}{M}\right)$ monotonically decreases and vanishes at extremality. 

It is instructive to compare the accretion rate in Eq. \eqref{constraint rades} with the Eddington accretion rate
$ \dot M_{Edd} = 4\pi M / (\bar \kappa_{es} \eta)$, where $\eta$ is the radiative efficiency defined in Eq. \eqref{radiative efficiency}. We find the reduced accretion rate
\be\label{MMEdd}
\dot m \equiv \frac{\dot M}{\dot M_{Edd}} = \frac{\eta}{\alpha} f \left(\frac{a}{M}\right).
\ee
Since the $\alpha$ parameter is usually hard to estimate, it is useful to solve the relation \eqref{MMEdd} for $\alpha$ in terms of $\dot m$ using Eq. \eqref{radiative efficiency}. The accretion rate is then determined from Eq. \eqref{constraint rades} and we obtain
\be\label{solalpha}
\alpha = \frac{\eta_0 f }{\dot m - h_0 f g } ,\qquad \dot M = \frac{4\pi M }{\eta_0 \bar \kappa_{es}}( \dot m - h_0 fg ).
\ee
The free parameters of the model where the ISCO lies in a radiation-dominated region can be finally taken to be $(M,a,\dot m,h_0)$.

In the phase diagrams of both typical stellar-mass and supermassive black holes displayed in Figs \ref{regions} and \ref{regions2}, the gas pressure dominates at the ISCO for all standard spins. However, for sufficiently high spins, radiation pressure dominates as we will now show. Let us first discuss configurations where the ISCO lies in the Region [Gas-es] for standard spins such as the case studied in Fig.~1. If one (wrongly) assumes that the ISCO lies in Region [Gas-es] for very high spins, one deduces from Eqs. \eqref{consistency gases1} and \eqref{h0 gases} that the ratio of pressures at the ISCO is 
\bea
\frac{p_{rad}}{p_{gas}}\Bigg|_{0} = \frac{\alpha \kappa_{es}}{2\sqrt{2}\pi} \frac{\dot M}{M} \frac{\sqrt{\mathcal D_0}}{\mathcal C_0 \sqrt{\mathcal R}_0 x_0^2} = 0.27 \frac{\alpha \dot M_{\star}}{M_\star} \sigma^{-2/3}+\mathcal{O}(\sigma),
\eea
where in the last step we took the near-extremal scaling $a/M=\sqrt{1-\sigma^2}$. For $\sigma \ll 1$, one obtains that radiation pressure will instead dominate. In the example of $M_\star = \dot M_\star = 1$ and $\alpha = 0.2$, the transition occurs (in the sense that $p_{rad}=p_{gas}$) at $a/M = 0.99996$ which is much above the Thorne bound of $0.998$ (\cite{1974ApJ...191..507T}) and therefore much probably unrealistic. 

If instead the ISCO lies in the Region [Gas-ff] for standard spins such as the case studied in Fig. \ref{regions2}, one deduces from Eqs. \eqref{consistency gasff1} and \eqref{h0B3} that the ratio of pressures at the ISCO is 
\be
\frac{p_{rad}}{p_{gas}}\Bigg|_0 = 0.02 \frac{\alpha^{8/17}\dot{M}_\star^{7/17}}{M_\star^{6/17}} \sigma^{-14/51}+\mathcal{O}\left(\sigma^{19/51}\right),
\ee
which is $\gg 1$ for $\sigma \ll 1$. Again, radiation pressure dominates for sufficiently high spins and a new \emph{near-ISCO region} opens up. However, for typical parameters $M_\star = 10^7$, $\dot M_\star = 10^5$ and $\alpha=0.2$, the transition to the near region occurs at $a/M=1-10^{-18}$ which is unreasonably high to be realistic.

However, there are more interesting parameters to consider. Let us  take an accretion rate at $23 \%$ Eddington ($\dot m = 0.23$) as a model for the plateau state of GRS 1915+105 (\cite{2041-8205-775-2-L45}) with the spin estimate $J/M^2=0.98$ (\cite{Reid:2014ywa}). Assuming a radiation-dominated ISCO, we can derive the $\alpha$ parameter using Eq. \eqref{solalpha} after fixing an estimate for $h_0$. We checked that for any value $0 < h_0< 0.01$, the resulting values of $\alpha=0.43$ and $\dot{M}_{\star} = 16.5$ differ by $1\%$ or less. The continuous transition between the gas-pressure- and radiation-dominated phases occurs at $a/M=0.980 \pm  0.001$. For definiteness, we choose $h_{0}=0.002$, so that the transition exactly occurs at $a/M=0.98$.

\begin{figure}
\centering
\includegraphics[width=8.5cm]{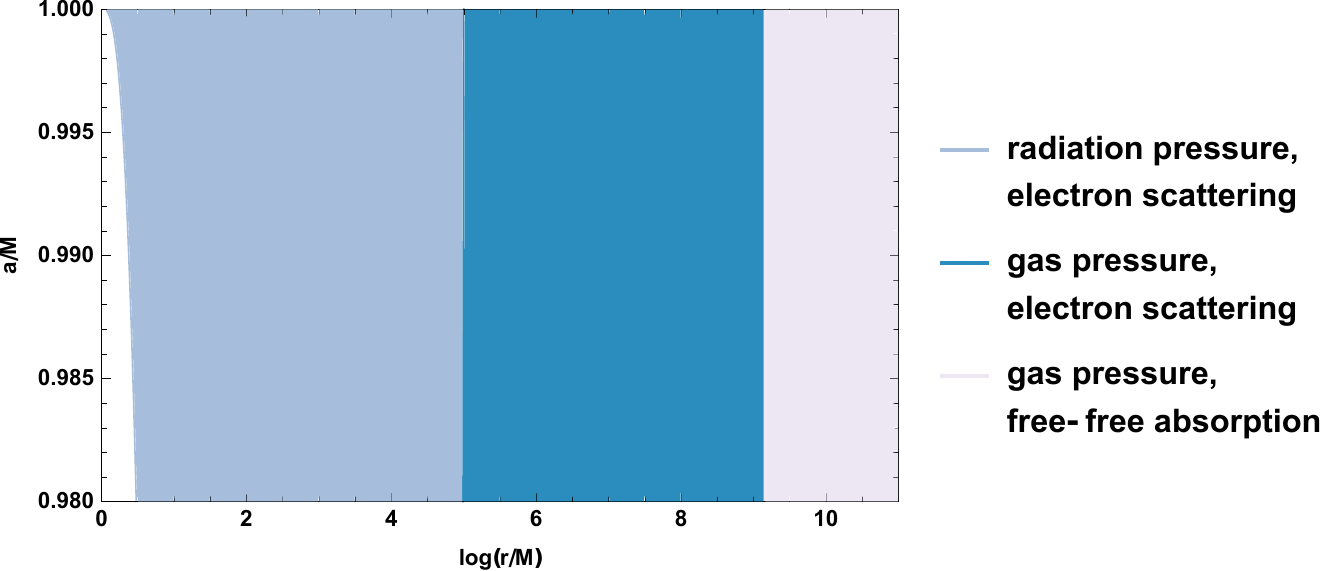}
\caption{Disk regions for fastly accreting highly spinning stellar-mass black hole with  $M_\star = 4$, accretion rate at $23\%$ of the Eddington limit and disk height at the ISCO $h_0 = 0.002$. The radiation-dominated region extends from the ISCO up to $r \sim 150 M$.}
\label{regions3}
\end{figure}

We conclude that GRS 1915+105 could be modelled by a high spin $a/M > 0.98$ and fastly accreting thin disk, which is radiation-dominated at the ISCO. As an example, we depict in Fig.~\ref{regions3} the phase diagram for the parameters $M_\star = 4$, $\dot m =0.23$, $h_0 =0.002$ and spins $0.98 < a/M < 1$. The \emph{inner region} is radiation-dominated. At finite radius away from the ISCO, there is a transition to a \emph{middle region}, which is gas pressure-dominated, but whose opacity is still dominated by electron scattering. There is also an \emph{outer region} further away, where the main mechanism for opacity is free-free absorption.

\section{Near-horizon near-extremal solution}
\label{NHEK sol}

In the near-extremal regime, a new spacetime region, the NHEK region, opens up as explained in Section \ref{sec:cr}, which is characterized by conformal symmetry. Since the radiation-dominated solution is the only one relevant in the limit of extremely high spins, we will perform the near-extremal near-horizon limit of the solution \eqref{sol rades} in order to exhibit its conformal  properties. 

We perform the scaling \eqref{NHEKs} with $a/M=\sqrt{1-\bar\sigma^2 \lambda^3}$ and let $\lambda \rightarrow 0$. The ISCO is located at $R_{0}=2^{1/3} \bar{\sigma}^{2/3}$. The accretion rate $\dot M$ (in terms of asymptotic time) is constrained as in Eq. \eqref{constraint rades} which implies the scaling $\dot M \sim \lambda$. In terms of near-horizon time $T$, the accretion rate $M^{\prime}$ is finite,
\be \label{constraint rades NHEK}
M^{\prime} = \frac{\p t}{\p T} \dot{M} = \frac{2M}{\lambda}\dot{M} = \pi \sqrt{6(7-\sqrt{3})}\frac{M^2}{\alpha \bar{\kappa}_{es}} R_{0}.
\ee
Therefore, contrary to the asymptotically flat observer, the NHEK observer measures a finite non-zero accretion rate $M^{\prime}$. The ratio $M^{\prime}/R_0$ is fixed by the parameters of the model. 

%Since $R_0$ can be arbitrarily scaled because of the dilatation operator $H_0$, the accretion rate $M^{\prime}$ depends where we fix the ISCO in NHEK.

The near-horizon behaviour of the solution \eqref{sol rades} can be obtained by performing the near-extremal near-horizon scaling and trading $\dot M$ for $M^\prime$. We get the following expressions at leading order ($O(\lambda^0)$): 
\begin{subequations} \label{NHEK sol rades}
	\begin{align}
		F &= \frac{7-\sqrt{3}}{4} \frac{h_0}{M \bar{\kappa}_{es}} \left(\frac{R_{0}}{R}\right)^2 \label{F scaling}\\
		&= \left(2.0 \times 10^{26} erg/(cm^2 sec)\right)\left(h_0 M_{\star}^{-1}\right)\left(\frac{R_{0}}{R}\right)^2, \nonumber\\
		\Sigma &= \frac{3}{2}\frac{1}{\alpha h_0 \bar{\kappa}_{es}} \left(\frac{R}{R_{0}}\right)^2 \\
		&= \left(3.75 g/cm^2\right)\left(\alpha^{-1} h_{0}^{-1}\right) \left(\frac{R}{R_{0}}\right)^2, \nonumber\\
		%W &= M F \\
		%&= \left(1.48 \times 10^{22} dyn/cm\right)\left(h_0\right)\left(\frac{R_{0}}{R}\right)^2, \nonumber\\
		h &=h_{0} \left(\frac{R_{0}}{R}\right)^2, \\
		u^{R} &= -\sqrt{\frac{7 - \sqrt{3}}{6}} \frac{h_0 R_0}{M} \left(\frac{R_{0}}{R}\right)^2 \\ 
		&= \left(-6.3 \times 10^{4} /sec\right)\left(h_0 R_{0} M_{\star}^{-1}\right) \left(\frac{R_{0}}{R}\right)^2, \nonumber \\
		p &= \frac{7 - \sqrt{3}}{8} \frac{1}{\alpha M \bar{\kappa}_{es}}  \\
		&=  \left(3.4 \times 10^{15} dyn/cm^2 \right)\left(\alpha^{-1} M_{\star}^{-1}\right), \nonumber\\
		\rho &= \frac{3}{4}\frac{1}{M \alpha h_0^2 \bar{\kappa}_{es}} \left(\frac{R}{R_0}\right)^4 \\
		&= \left(4.21 \times 10^{-6} g/cm^3 \right)\left(\alpha^{-1} h_{0}^{-2}M_{\star}^{-1}\right)  \left(\frac{R}{R_0}\right)^4, \nonumber\\
		T &= \left(\frac{3(7 - \sqrt{3})}{8\alpha b M \bar{\kappa}_{es}}\right)^{1/4} = \left(3.39 \times 10^{7} K \right)\left(\alpha^{-1/4} M_{\star}^{-1/4}\right). \label{Tnear}
	\end{align}
\end{subequations}
All these quantities are defined for $R \geq R_0$. Note that all quantities, except $p$ and $T$, depend on the ratio
\be
\frac{R_{0}}{R} = \frac{\left[2\left(1-(a/M)^2\right)\right]^{1/3}}{x^2-1},
\ee
which is independent of the choice of the constant $\bar{\sigma}$. The only exception is the radial component of the four-velocity, which has an additional power of $R_0$, and therefore depends on the position of the ISCO. The radial scaling exponents reproduce the table of conformal weights in Eq. \eqref{cw}, announced in Section \ref{sec:cr}. Note one unusual property of the self-similar solution: starting from $h_0$ at the ISCO the disk height $h$ \emph{decreases} with the radius; it increases again outside of the range of validity of the self-similar solution.

\begin{figure}
\centering
\includegraphics[width=8.5cm]{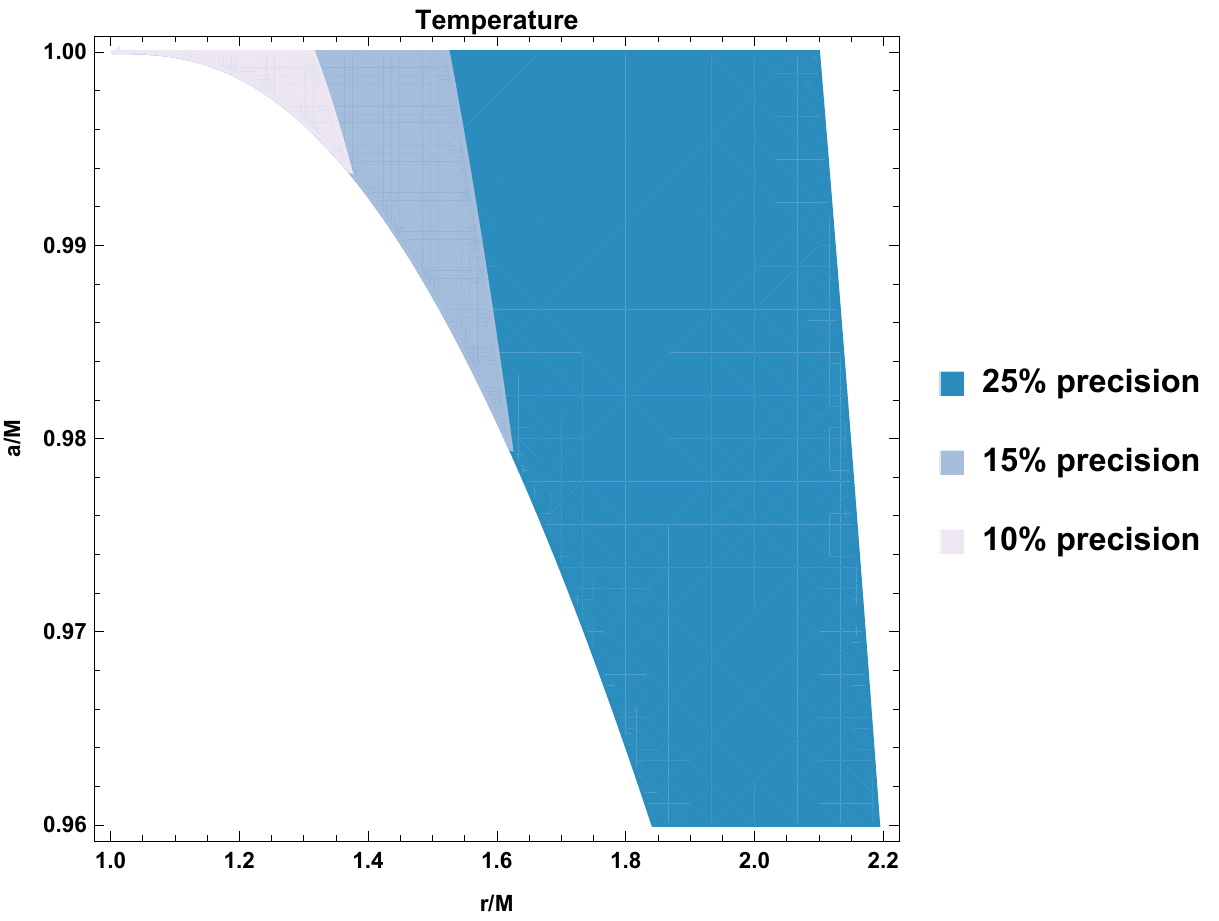}
\caption{Relative precision of the self-similar critical temperature with respect to the actual temperature profile in the radiation-dominated region around the ISCO as a function of the spin. The relative precision is independent of other parameters of the model.}
\label{Prec1}
\end{figure}

The self-similar solution is an approximate solution of the disk around the ISCO. It is important to discuss its range of validity. The critical temperature does not depend upon the accretion rate $\dot M$ or the height of the disk $h_0$. Since $\alpha$ or $M$ are overall factors of the temperature profile \eqref{T inner}, the relative error between the actual temperature profile and the constant critical temperature only depends upon the radius and the spin. We find that for near-extremal spins $0.96 \leq a/M \leq 1$, the actual temperature profile deviates from the critical temperature by less than $25\%$ only in the range $r_{0} \leq r \leq 2.1-2.2 M$, where the upper bound is nearly independent of the spin. This very limited \emph{near-ISCO region} is the region where the disk is approximately described by the self-similar solution. The region is biggest when the ISCO approaches $M$, which occurs closest to extremality. If a higher precision is required, the region of validity shrinks accordingly. We plot in Fig. \ref{Prec1} the range of validity of the temperature of the self-similar solution with 25$\%$, $15\%$ and $10\%$ relative precision.

\begin{figure}
\centering
\includegraphics[width=8.5cm]{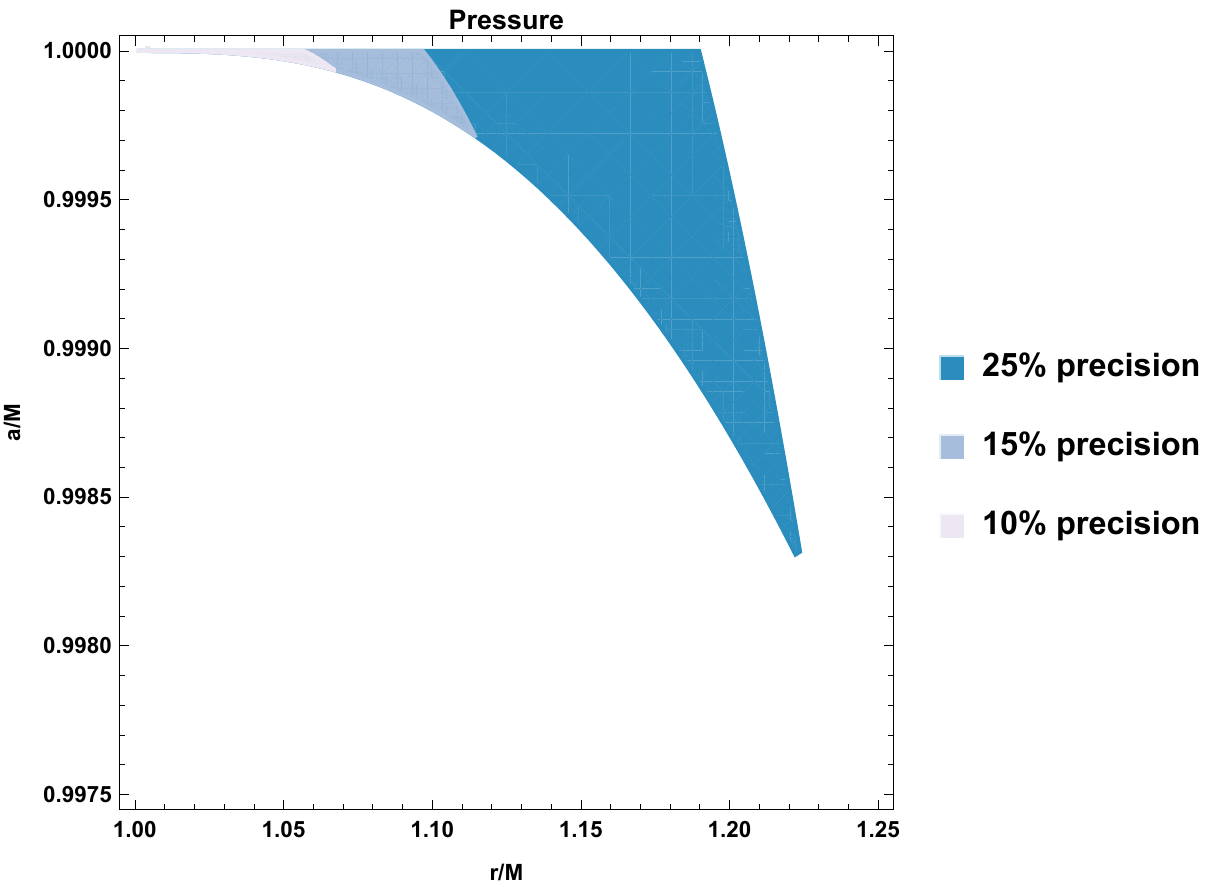}
\caption{Relative precision of the self-similar critical pressure with respect to the actual pressure profile in the radiation-dominated region around the ISCO as a function of the spin. The relative precision is independent of other parameters of the model.}
\label{Prec2}
\end{figure}

\begin{figure}
\centering
\includegraphics[width=8.5cm]{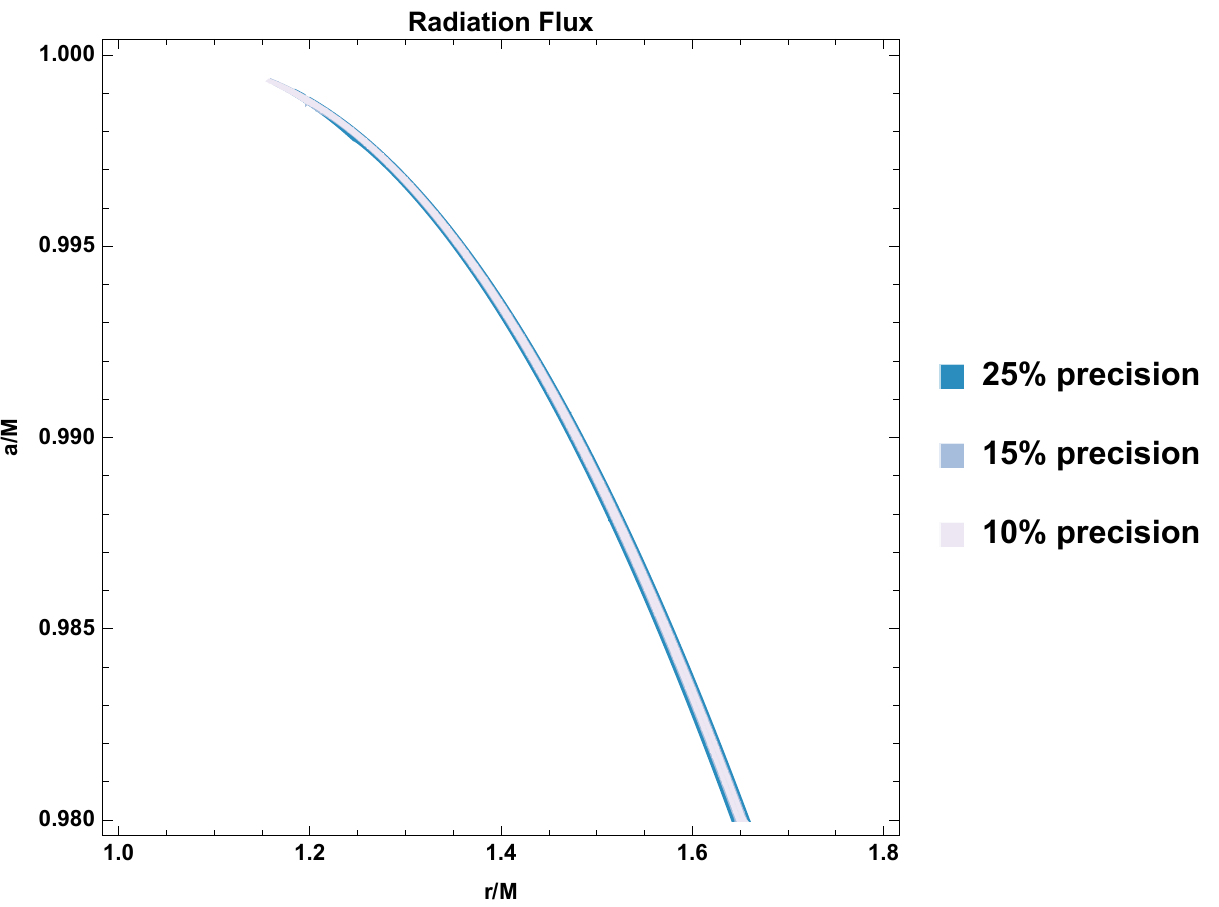}
\caption{Relative precision of the self-similar critical radiation flux with respect to the actual radiation flux profile in the radiation-dominated region around the ISCO as a function of the spin for $h_{0}= 0.002$, $\dot m =0.23$. The auxiliary parameter $\alpha$ is fixed through Eq. \eqref{solalpha}. The relative precision is independent of the mass, because it factors out upon substituting the accretion rate \eqref{solalpha} into Eq. \eqref{F rades} and taking the ratio with Eq. \eqref{F scaling}.} 
\label{Prec3}
\end{figure}

Other physical quantities can be analysed similarly. Unfortunately, the relative precision of the pressure requires a spin higher than the Thorne bound $a/M=0.998$ and a narrower region around the ISCO, as plotted in Fig. \ref{Prec2}. Another important physical quantity is the radiation flux $F$. Fig. \ref{Prec3} shows that the range of validity of the self-similar solution is limited to a narrow region around the ISCO. The physical relevance of the self-similar solution \eqref{NHEK sol rades} is therefore uncertain.

In conclusion, we found a first example of critical accretion around a near-extremal black hole, which admits a scale invariance in a region close to the ISCO. More elaborated models are required to find more realistic solutions. 

\vspace{-0.2cm}
\section*{Acknowledgements}
We are grateful to A. Lupsasca and A. Strominger for their comments and especially A. Lupsasca for pointing out several typos. We thank our referee, M. Abramowicz, for constructive remarks. G.C. and R.O. acknowledge  the  current  support  of  the  ERC  Starting  Grant  No.   335146  ``HoloBHC''. G.C. is a Research Associate of the Fonds de la Recherche Scientifique F.R.S.-FNRS (Belgium).

\appendix

\section{Kinematics around Kerr}\label{appKerr}

In this appendix we adopt geometric units $G = c = 1$. 

The Kerr line element \cite{Bardeen:1972fi} in BL coordinates $(t, r, \theta, \phi)$ is given by
\be \label{Kerr}
ds^2 =-\frac{\Sigma \Delta}{A} dt^2 + \sin^2(\theta) \frac{A}{\Sigma}(d\phi - \omega dt)^2 + \frac{\Sigma}{\Delta}dr^2 + \Sigma d\theta^2,
\ee
where the metric functions read
\begin{subequations}
\begin{align}
\Delta(r) &= r^2-2Mr +a^2, \\
\Sigma(r,\theta) &= r^2+a^2\cos^2(\theta),\\
A(r, \theta) &= (r^2+a^2)^2-a^2\Delta\sin^2(\theta),\\
\omega(r, \theta) &= \frac{2 M a r }{A}.
\end{align}
\end{subequations}
Here, $M$ is the mass and $a = J/M$ is the specific angular momentum of the Kerr black hole, which we assume to be non-negative. The event horizon is located at $r_{+} = M + (M^2 - a^{2})^{1/2}$. The Kerr black hole spacetime is stationary and axisymmetric. The Killing vectors corresponding to time and axial symmetry are, respectively, $\eta = \delta^{\mu}_{\; t} \p_{\mu}$ and $\xi = \delta^{\mu}_{\; \phi} \p_{\mu}$.

In order to describe the region near the equatorial plane $\theta = \pi / 2$, we introduce the cylindrical coordinate $z = r \cos(\theta)$ and neglect corrections of order $(z/r)^2$. The resulting line element in the near-equatorial region of coordinates $(t, r, z, \phi)$ is given by
\be \label{Kerr NE}
ds^2 = -\frac{r^2 \Delta}{A}dt^2 + \frac{A}{r^2}(d\phi - \omega dt)^2 + \frac{r^2}{\Delta}dr^2 + dz^2 + \mathcal{O}\left(\frac{z}{r}\right)^2.
\ee
Following \cite{25334}, we define the dimensionless radial coordinate $x$ and the spin parameter $a_s$, respectively, by
\be \label{dimless}
x \equiv \left(\frac{r}{M}\right)^{\frac{1}{2}}, \qquad a_s \equiv \frac{a}{M}
\ee
Thus, at fixed spin parameter in the range $a_{s} \in [0,1]$, $x \geq x_{+} =\left[1+ (1-a_s^2)^{1/2}\right]^{1/2}$ describes the Kerr spacetime outside the event horizon.
The ISCO is located at
\begin{subequations} \label{ISCO}
\begin{align}
x_0 &= \left\{3 + Z_2 -\left[(3-Z_1)(3+Z_1+2Z_2)\right]^{1/2}   \right\}^{1/2},\\
%\end{align}
%\end{subequations}
%\begin{subequations}
%\begin{align}
Z_1 &=1 + \left(1-a_s^2\right)^{1/3}\left[\left(1+a_s\right)^{1/3}+\left(1-a_s\right)^{1/3}\right],\\
Z_2 &= \left(3a_s^2+Z_1^2\right)^{1/2}.
\end{align}
\end{subequations}
Following the tradition, we define the dimensionless functions\footnote{Note the typo in the exponent of $a_s$ in Eq. (A4c) of \cite{Penna:2011rw}.}:
\begin{subequations} \label{AtoG}
\begin{align} 
\mathcal{A} &= 1+\frac{a_s^2}{x^4} +\frac{2a_s^2}{x^6}, \\
\mathcal{B} &= 1+ \frac{a_s}{x^3},\\
\mathcal{C} &= 1-\frac{3}{x^2}+\frac{2a_s}{x^3},\\
\mathcal{D} &= 1-\frac{2}{x^2}+\frac{a_s^2}{x^4},\\
\mathcal{F} &= 1-\frac{2a_s}{x^3}+\frac{a_s^2}{x^4},\\
\mathcal{G} &= 1-\frac{2}{x^2}+\frac{a_s}{x^3},\\
\mathcal{R} &= \mathcal{C}^{-1}\mathcal{F}^2 - a_s^2 x^{-2}\left(\mathcal{C}^{-1/2}\mathcal{G}-1\right).
\end{align}
\end{subequations}
All functions \eqref{AtoG} go to unity far from the black hole $x \gg 1$.
The functions $\mathcal{A}$, $\mathcal{B}$ and $\mathcal{R}$ are monotonically decreasing, while the functions $\mathcal{C}$, $\mathcal{D}$, $\mathcal{F}$ and $\mathcal{G}$ are monotonically increasing in their domains of definition. The radial coordinates where $\mathcal{C} =0 $ are given by 
\begin{subequations}
\begin{align}
x_1 &= 2 \cos \left[ \frac{1}{3}\arccos(a_s) - \frac{\pi}{3} \right] ,\\
x_2 &= 2 \cos \left[ \frac{1}{3}\arccos(a_s) + \frac{\pi}{3} \right] ,\\
x_3 &= -2 \cos \left[ \frac{1}{3}\arccos(a_s)  \right],
\end{align}
\end{subequations}
with $x_1 \geq x_2 \geq x_3$. The last photon orbit is given by $x=x_1$ and is placed between the ISCO and the event horizon, $x_0 \geq x_1 \geq x_+$. The function $\mathcal{D}$ is positive outside the event horizon.

\subsection{Circular equatorial geodesics} \label{CEGs}

Circular equatorial geodesics have a four-velocity of the form
\begin{subequations}
\begin{align}
u = u^{t}\p_t + u^{\phi}\p_{\phi} = u^{t}(\eta + \Omega \xi).
\end{align}
\end{subequations}
Here, we list the kinematic quantities characterizing the geodesic motion:
\begin{subequations} \label{kin quantities1}
\begin{align}
u^{t} &= \mathcal{B}\mathcal{C}^{-1/2},\\
u^{\phi} &= M^{-1} x^{-3} \mathcal{C}^{-1/2},\\
E &= \eta^{\mu} u_{\mu} = \mathcal{C}^{-1/2} \mathcal{G}, \\
L &= \xi^{\mu} u_{\mu} = M x  \mathcal{C}^{-1/2} \mathcal{F},\\
\Omega &= u^{\phi}/u^{t} = M^{-1} x^{-3} \mathcal{B}^{-1},\\
\mathcal{L}^{2}_{\star} &= L^{2} - a^2(E - 1) = M^2 x^2 \mathcal{R},
\end{align}
\end{subequations}
respectively, the time component $u^{t}$, the azimuthal component $u^{\phi}$, the conserved specific energy $E$, the conserved specific angular momentum $L$, the corotating Keplerian angular velocity $\Omega$ with respect to a stationary observer and a conserved quantity along the geodesic motion $\mathcal{L}^{2}_{\star}$, which appears in the thin disk vertical equation.

It is useful to introduce the function $\mathcal{P}$ defined by
\begin{align} \label{P}
&\mathcal{P}  \equiv \mathcal{P}_0+ \frac{1}{M} \int_{x_0}^{x} (E- \Omega L) L_{,x^{\prime}} dx^{\prime} \nonumber\\
&= \mathcal{P}_0+ x - x_0 - \frac{3}{2} a_s \ln\left(\frac{x}{x_0}\right)\nonumber\\
&\hspace{1cm} - \sum_{i=1}^{3}\frac{3(x_i-a_s)^2}{x_i(x_i-x_{i+1})(x_i-x_{i+2})}\ln \left(\frac{x-x_i}{x_0-x_i} \right).
\end{align}
Here, $\mathcal{P}_0$  is a real constant to be fixed by imposing a boundary condition, as explained in Sec. \ref{ISCOBC}.
In the above sum, we identify $x_i = x_{i+3}$. The function $\mathcal Q$ of \cite{25341} can be expressed as $\mathcal Q = x^{-1}\mathcal B \mathcal C^{-1/2}\mathcal P$ but we find more economical and natural to use $\mathcal P$. 

\subsection{Observer frames}
We define three privileged observers.
\begin{enumerate}
\item The stationary observer frame with four-velocity 
\be
u_{stat} = (- \eta^{\nu}\eta_{\nu})^{-1/2} \eta,
\ee 
which is locally rotating in the sense that $L_{stat} = \xi_{\mu} u_{stat}^{\mu} \neq 0$. 

\item The locally nonrotating frame (LNRF), sometimes also called local inertial frame or zero angular momentum observer (ZAMO), has four-velocity
\be
u_{LNRF} = ( \omega^{2}\xi^{\mu}\xi_{\mu} -\eta^{\mu}\eta_{\mu})^{-1/2} ( \eta + \omega \xi), 
\ee
with $\omega^{2}\xi^{\mu}\xi_{\mu} -\eta^{\mu} \eta_{\mu}= \mathcal{A}^{-1}\mathcal{D}$.
It has $L_{LNRF} =  \xi_{\mu} u_{LNRF}^{\mu} = 0$.
With respect to the LNRF observer, circular equatorial geodesics have four-velocity components $u^{(a)} = u^{\mu} e_{\mu}^{\; (a)}$, where $e_{\mu}^{\; (a)}$ is the orthonormal tetrad attached to the LNRF, whose expression can be found in Eq. (3.2) of \cite{Bardeen:1972fi}. The only two non-vanishing components are
\begin{align}
u^{(t)} &=  \frac{r \Delta^{1/2}}{A^{1/2}} u^{t}= \mathcal{A}^{-1/2} \mathcal{B}\mathcal{C}^{-1/2}\mathcal{D}^{1/2},\\
u^{(\phi)} &=\frac{A^{1/2}}{r}(\Omega -\omega) u^{t} \\ \nonumber
&=x^{-1} \mathcal{A}^{1/2}  \mathcal{C}^{-1/2} - 2 a_s x^{-4} \mathcal{A}^{-1/2}  \mathcal{B} \mathcal{C}^{-1/2}.
\end{align}
Therefore, the only nonvanishing component of the three-velocity relative to the LNRF is
\be 
\mathcal{V}^{(\phi)} = \frac{u^{(\phi)}}{u^{(t)}} =\frac{A}{r^2 \Delta^{1/2}}(\Omega - \omega) =  x^{-1} \mathcal{B}^{-1}\mathcal{D}^{-1/2} \mathcal{F}.
\ee
We call $\mathcal{V}^{(\phi)} = \tilde{R}(\Omega_k - \omega)$ the linear velocity with respect to the LNRF , with $ \tilde{R} = A/(r^2 \Delta^{1/2})$ being the gyration radius and $\gamma = u^{(t)}$ the Lorentz gamma factor corresponding to this linear velocity. In other words, the four-velocity of circular equatorial geodesics as measured by an LNRF observer is $u^{\mu} = \gamma (u_{LNRF}^{\mu} + \mathcal{V}^{(\phi)}e_{(\phi)}^{\mu})$.
\item The local rest frame (LRF) of the particle in the circular equatorial orbit, whose orthonormal tetrad is defined in Eq. (5.4.5a) of \cite{25334}.
In this frame, the shear tensor $\sigma_{\hat{\mu}\hat{\nu}}$ of circular equatorial geodesics has only one non-vanishing component 
\be \label{shear tensor}
\sigma_{\hat{r}\hat{\phi}} =\sigma_{\hat{\phi}\hat{r}} = \frac{1}{2} \frac{A}{r^3} \gamma^2 \Omega_{,r} = -\frac{3}{4}M^{-1} x^{-3}\mathcal{C}^{-1}\mathcal{D}.
\ee
\end{enumerate}

\onecolumn
\section{Local solutions} \label{Solutions}

We have eight equations, \eqref{surface density}, \eqref{int shear tensor}, \eqref{re1}, \eqref{re2}, \eqref{re3}, \eqref{vertical equation1},  \eqref{EOS}, \eqref{vertical equation2} and eight unknown functions $F$, $\Sigma$, $W$, $h$, $u^{r}$, $p$, $\rho$, $T$ of the radial coordinate. The system of equations is algebraic and admits  a single solution upon imposing the physical conditions that $T>0$, $p>0$. The solution depends upon four free parameters. The solution can be patched by local solutions where either the gas pressure or radiation pressure dominates, and opacity is either dominated by electron scattering or free-free absorption. The three relevant local solutions are detailed below and are denoted as 
\begin{enumerate}
\item[\bf Gas-es:] Gas pressure-electron scattering dominated: $p=p^{(gas)}$ and $\bar{\kappa} = \bar{\kappa}_{es}$;
\item[\bf Rad-es:] Radiation pressure-electron scattering dominated solution: $p=p^{(rad)}$ and $\bar{\kappa} = \bar{\kappa}_{es}$;
\item[\bf Gas-ff:] Gas pressure-free free absorption dominated solution: $p=p^{(gas)}$ and $\bar{\kappa} = \bar{\kappa}_{ff}$.
%\item[-] B4 where $p=p^{(rad)}$ and $\bar{\kappa} = \bar{\kappa}_{ff}$;
\end{enumerate} 
The numerical values in cgs units of all physical constants used in the solutions are
\bea
G = 6.67 \times 10^{-8} cm^3/(sec^2 g), \quad c = 3.00 \times 10^{10} cm/sec, \quad k_{B} = 1.38  \times 10^{-16} erg/K, \nonumber \\
m_{p} = 1.67  \times 10^{-24} g, \quad  b = 7.56 \times 10^{-15} erg/(cm^3 K^4), \quad M_{\astrosun} = 1.99  \times 10^{33} g.
\eea
The Novikov-Thorne local solutions (with typos fixed) can be recovered by substituting $x \rightarrow \sqrt{r_*}$, $\mathcal P \rightarrow r_*^{1/2} \mathcal B^{-1}\mathcal C^{1/2} \mathcal Q$ and using the boundary condition at the ISCO $\mathcal Q(r_0)=0$.

\begin{enumerate}
\item[{\bf [Gas-es]}]{\bf Gas pressure-electron scattering dominated solution}\\
In this region, the gas pressure $p^{(gas)}$ is predominant with respect to the radiation pressure $p^{(rad)}$ and the electron scattering contributes mainly to the opacity of the disk. 
\begin{subequations} \label{sol gases}
\begin{align}
F 				&= \left(5.5 \times 10^{25} erg/(cm^2 sec)\right) \left(M_{\star}^{-2} \dot{M}_{\star}\right) x^{-7} \mathcal{C}^{-1} \mathcal{P},\\
\Sigma   &=\left(5.0 \times 10^4 g/cm^2 \right) \left(\alpha^{-4/5} M_{\star}^{-2/5} \dot{M}_{\star}^{3/5}\right) x^{-9/5} \mathcal{C}^{1/5} \mathcal{D}^{-4/5} \mathcal{P}^{3/5},\\
W				&=\left(1.1 \times 10^{21} dyn/cm\right) \left(M_{\star}^{-1} \dot{M}_{\star}\right) x^{-4} \mathcal{D}^{-1} \mathcal{P},\\
h				&=\left(7.0 \times 10^{-3}\right) \left(\alpha^{-1/10} M_{\star}^{-3/10} \dot{M}_{\star}^{1/5}\right) x^{-1/10} \mathcal{C}^{-1/10} \mathcal{D}^{-1/10} \mathcal{R}^{-1/2} \mathcal{P}^{1/5}, \label{h gases}\\
u^{r}		&=\left(-7.3 \times 10^5  cm/sec \right) \left(\alpha^{4/5} M_{\star}^{-3/5} \dot{M}_{\star}^{2/5}\right) x^{-1/5} \mathcal{C}^{-1/5} \mathcal{D}^{4/5} \mathcal{P}^{-3/5},\\
p^{(gas)}&=\left(1.8 \times 10^{17} dyn/cm^2\right) \left(\alpha^{-9/10} M_{\star}^{-17/10} \dot{M}_{\star}^{4/5}\right) x^{-59/10} \mathcal{C}^{1/10} \mathcal{D}^{-9/10} \mathcal{R}^{1/2} \mathcal{P}^{4/5},\\
\rho			&=\left(8.1 g/cm^3\right) \left(\alpha^{-7/10} M_{\star}^{-11/10} \dot{M}_{\star}^{2/5}\right) x^{-37/10} \mathcal{C}^{3/10} \mathcal{D}^{-7/10} \mathcal{R}^{1/2} \mathcal{P}^{2/5},\\
T				&=\left(2.6 \times 10^8 K\right) \left(\alpha^{-1/5} M_{\star}^{-3/5} \dot{M}_{\star}^{2/5}\right) x^{-11/5} \mathcal{C}^{-1/5} \mathcal{D}^{-1/5} \mathcal{P}^{2/5}.
\end{align}
\end{subequations}

For consistency, this solution is valid where $h\ll1$, $u^r \ll1$ and 
\begin{subequations} \label{consistency gases}
\begin{align}  \label{consistency gases1}
\frac{p^{(rad)}}{p^{(gas)}}			&=\left(69.\right) \left(\alpha^{1/10} M_{\star}^{-7/10} \dot{M}_{\star}^{4/5}\right) x^{-29/10} \mathcal{C}^{-9/10} \mathcal{D}^{1/10} \mathcal{R}^{-1/2} \mathcal{P}^{4/5} \ll 1,\\
\frac{\bar{\kappa}_{ff}}{\bar{\kappa}_{es}}			&=\left(4.4 \times 10^{-6}\right) \left(M_{\star} \dot{M}_{\star}^{-1}\right) x^{4} \mathcal{C} \mathcal{R}^{1/2} \mathcal{P}^{-1} \ll1. \label{consistency gases tau}
\end{align}
\end{subequations}

\item[{\bf [Rad-es]}]{\bf Radiation pressure-electron scattering dominated solution}\\
In this region, the radiation pressure $p=p^{(rad)}$ is dominant, but still the electrons scattering is the main mechanism for the opacity in the disk. 
\begin{subequations} \label{sol rades}
\begin{align}
F 				&= \left(5.5 \times 10^{25} erg/(cm^2 sec)\right) \left(M_{\star}^{-2} \dot{M}_{\star}\right) x^{-7} \mathcal{C}^{-1} \mathcal{P}, \label{F rades}\\
\Sigma   &=\left(10. g/cm^2 \right) \left(\alpha^{-1} M_{\star} \dot{M}_{\star}^{-1}\right) x^{4} \mathcal{C}^{2} \mathcal{D}^{-1} \mathcal{R} \mathcal{P}^{-1},\\
W				&=\left(1.1 \times 10^{21} dyn/cm\right) \left(M_{\star}^{-1} \dot{M}_{\star}\right) x^{-4} \mathcal{D}^{-1} \mathcal{P},\\
h				&=\left(0.5 \right) \left( M_{\star}^{-1} \dot{M}_{\star} \right) x^{-3} \mathcal{C}^{-1} \mathcal{R}^{-1} \mathcal{P}, \label{h rades}\\
u^{r}		&=\left(-3.5 \times 10^9  cm/sec \right) \left(\alpha M_{\star}^{-2} \dot{M}_{\star}^{2}\right) x^{-6} \mathcal{C}^{-2} \mathcal{D} \mathcal{R}^{-1} \mathcal{P},\\
p^{(rad)}&=\left(2.6 \times 10^{15} dyn/cm^2\right) \left(\alpha^{-1} M_{\star}^{-1} \right) x^{-3} \mathcal{C} \mathcal{D}^{-1} \mathcal{R},\\
\rho			&=\left(2.5 \times 10^{-5} g/cm^3\right) \left(\alpha^{-1} M_{\star} \dot{M}_{\star}^{-2}\right) x^{5} \mathcal{C}^{3} \mathcal{D}^{-1} \mathcal{R}^{2} \mathcal{P}^{-2},\\
T				&=\left(3.2 \times 10^7 K\right) \left(\alpha^{-1/4} M_{\star}^{-1/4} \right) x^{-3/4} \mathcal{C}^{1/4} \mathcal{D}^{-1/4} \mathcal{R}^{1/4}. \label{T inner}
\end{align}
\end{subequations}
For consistency, this solution is valid where $h\ll1$, $u^r \ll1$ and 
\begin{subequations} \label{consistency rades}
\begin{align}
\frac{p^{(gas)}}{p^{(rad)}}			&=\left(2.6 \times 10^{-5}\right) \left(\alpha^{-1/4} M_{\star}^{7/4} \dot{M}_{\star}^{-2}\right) x^{29/4} \mathcal{C}^{9/4} \mathcal{D}^{-1/4} \mathcal{R}^{5/4} \mathcal{P}^{-2} \ll 1,\\
\frac{\bar{\kappa}_{ff}}{\bar{\kappa}_{es}}			&=\left(2.2 \times 10^{-8}\right) \left( \alpha^{-1/8} M_{\star}^{15/8} \dot{M}_{\star}^{-2}\right) x^{61/8} \mathcal{C}^{17/8} \mathcal{D}^{-1/8} \mathcal{R}^{9/8} \mathcal{P}^{-2} \ll1.
\end{align}
\end{subequations}

\item[{\bf [Gas-ff]}]{\bf Gas pressure-free free absorption dominated solution}\\
In this region, the gas pressure $p=p^{(gas)}$ and the free-free term contribution in the opacity law are dominant. Then, we find
\begin{subequations} \label{sol gasff}
\begin{align}
F 				&= \left(5.5 \times 10^{25} erg/(cm^2 sec)\right) \left(M_{\star}^{-2} \dot{M}_{\star}\right) x^{-7} \mathcal{C}^{-1} \mathcal{P},\\
\Sigma   &=\left(1.70 \times 10^5 g/cm^2 \right) \left(\alpha^{-4/5} M_{\star}^{-1/2} \dot{M}_{\star}^{7/10}\right) x^{-11/5} \mathcal{C}^{1/10} \mathcal{D}^{-4/5} \mathcal{R}^{-1/20} \mathcal{P}^{7/10},\\
W				&=\left(1.1 \times 10^{21} dyn/cm\right) \left(M_{\star}^{-1} \dot{M}_{\star}\right) x^{-4} \mathcal{D}^{-1} \mathcal{P},\\
h				&=\left(3.8 \times 10^{-3}\right) \left(\alpha^{-1/10} M_{\star}^{-1/4} \dot{M}_{\star}^{3/20}\right) x^{1/10} \mathcal{C}^{-1/20} \mathcal{D}^{-1/10} \mathcal{R}^{-19/40} \mathcal{P}^{3/20},\label{hB3}\\
u^{r}		&=\left(-2.1 \times 10^5  cm/sec \right) \left(\alpha^{4/5} M_{\star}^{-1/2} \dot{M}_{\star}^{3/10}\right) x^{1/5} \mathcal{C}^{-1/10} \mathcal{D}^{4/5} \mathcal{R}^{1/20} \mathcal{P}^{-7/10},\\
p^{(gas)}&=\left(3.3 \times 10^{17} dyn/cm^2\right) \left(\alpha^{-9/10} M_{\star}^{-7/4} \dot{M}_{\star}^{17/20}\right) x^{-61/10} \mathcal{C}^{1/20} \mathcal{D}^{-9/10} \mathcal{R}^{19/40} \mathcal{P}^{17/20},\\
\rho			&=\left(51. g/cm^3\right) \left(\alpha^{-7/10} M_{\star}^{-5/4} \dot{M}_{\star}^{11/20}\right) x^{-43/10} \mathcal{C}^{3/20} \mathcal{D}^{-7/10} \mathcal{R}^{17/40} \mathcal{P}^{11/20},\\
T				&=\left(7.7 \times 10^7 K\right) \left(\alpha^{-1/5} M_{\star}^{-1/2} \dot{M}_{\star}^{3/10}\right) x^{-9/5} \mathcal{C}^{-1/10} \mathcal{D}^{-1/5} \mathcal{R}^{1/20} \mathcal{P}^{3/10}.
\end{align}
\end{subequations}
For consistency, this solution is valid where $h\ll1$, $u^r \ll1$ and 
\begin{subequations} \label{consistency gasff}
\begin{align}
\frac{p^{(rad)}}{p^{(gas)}}			&=\left(0.27\right) \left(\alpha^{1/10} M_{\star}^{-1/4} \dot{M}_{\star}^{7/20}\right) x^{-11/10} \mathcal{C}^{-9/20} \mathcal{D}^{1/10} \mathcal{R}^{-11/40} \mathcal{P}^{7/20} \ll 1, \label{consistency gasff1}\\
\frac{\bar{\kappa}_{es}}{\bar{\kappa}_{ff}}			&=\left( 4.8 \times 10^2  \right) \left(M_{\star}^{-1/2} \dot{M}_{\star}^{1/2} \right) x^{-2} \mathcal{C}^{-1/2} \mathcal{R}^{-1/4} \mathcal{P}^{1/2} \ll1.
\end{align}
\end{subequations}

\end{enumerate}

%\twocolumn

%%%%%%%%%%%%%%%%%%%%%%%%%%%%%%%%%%%%%%%%%%%%%%%%%%



\bibliographystyle{mnras}
\bibliography{refsDisks} % if your bibtex file is called example.bib

%%%%%%%%%%%%%%%%%%%%%%%%%%%%%%%%%%%%%%%%%%%%%%%%%%

% Don't change these lines
\bsp	% typesetting comment
\label{lastpage}
\end{document}